\documentclass[fleqn,prl,reprint,aps,citesort]{revtex4-1}
\usepackage{graphicx,amsmath,amssymb,booktabs,subfigure,hyperref,pifont,color}

\newcommand{\del}{\partial}

\renewcommand{\v}[1]{\mathbf{#1}}

\newcommand{\ave}[1]{\langle {#1} \rangle}

\begin{document}
\title{Perfect fluid flow from granular jet impact}
\author{Jake Ellowitz, Nicholas Guttenberg, Wendy W.\ Zhang}
\affiliation{The James Franck Institute at the University of Chicago, Chicago IL 60637}
\date{\today}

\begin{abstract}Experiments on the impact of a densely-packed jet of non-cohesive grains onto a fixed target show that the impact produces an ejecta sheet comprised of particles in collimated motion. The ejecta sheet leaves the target at a well-defined angle whose value agrees quantitatively with the sheet angle produced by water jet impact. Motivated by these experiments, we examine the idealized problem of dense granular jet impact onto a frictionless target in two dimensions. Numerical results for the velocity and pressure fields within the granular jet agree quantitatively with predictions from an exact solution for 2D perfect-fluid impact. This correspondence demonstrates that the continuum limit controlling the coherent collective motion in dense granular impact is Euler flow. 
\end{abstract}
\maketitle

Impact is a powerful tool in physics. We use scattering techniques to deduce the structure of matter by analyzing the ejecta distribution produced by the impact of a highly collimated beam against a target. Here we examine a macroscopic version of beam impact: the collision of a jet of macroscopic grains against a fixed target. We find that to leading order, the impact of a dense stream of grains is described by Euler flow. That is, the granular dynamics simplifies greatly and is governed primarily by incompressibility and momentum conservation.

\begin{figure}
\includegraphics[width=1.0\columnwidth]{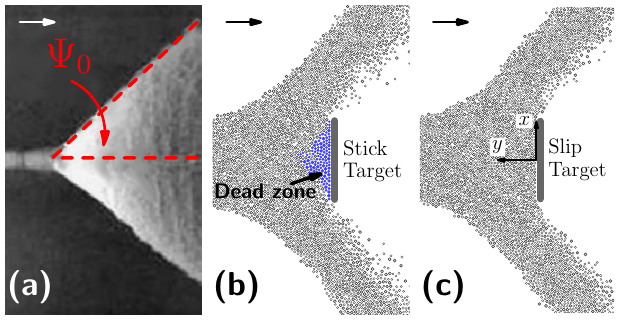}
\caption{\label{fig:ejected_sheet} (Color online)
{\bf (a)} Snapshot of the axisymmetric, hollow ejecta sheet formed by dense non-cohesive granular jet impact. The sheet leaves the target with an angle $\Psi_0$ (from \cite{cheng_2007}). 
{\bf (b)} Snapshot of a 2D simulation of granular impact onto a stick target: once a grain touches the target, it remains motionless for the rest of the simulation. Here, $D_{\rm grain} = D_{\rm tar}/25$. 
A dead zone marked by the blue grains is present. The dead zone is distinct from a shock because the density is continuous across the dead zone surface.
Smaller $D_{\rm grain}/D_{\rm tar}$ increases the coherence of the ejecta. 
{\bf (c)} Snapshot of a 2D simulation of granular jet impact onto a frictionless, or slip, target, with $D_{\rm grain} = D_{\rm tar}/25$.
In 2D, the collimated ejecta form two outgoing jets instead of the conical shape we see in 3D.
The collimated ejecta forms in 2D and 3D, and with or without the interior dead zone.}
\end{figure}

Our work is motivated by experiments showing that the collision of a dense granular jet with a fixed target produces a liquid-like dynamical response~\cite{cheng_2007}. 
At low volume fractions, impact of a granular stream onto a fixed target produces a shock which quantitatively agree with the Mach cone produced by supersonic gas flow~\cite{swinney02,kellay09, wassgren_shocks}. At intermediate volume fractions, experiments and simulations show coexistence of a shock-like front coexisting with a dead zone~\cite{kellay01,knoll07,hogg05,gray03}.
In contrast, when the jet is densely-packed, the amount of scatter is significantly reduced. Instead, the angles of recoil are narrowly distributed about a mean value $\Psi_0$ (Fig.~\ref{fig:ejected_sheet}\textbf{(a)}). As a result, the granular jet is ejected away from the target in a thin axisymmetric sheet forming the shape of a hollow cone, reminiscent of the ``water bell''~\cite{clanet_2001}.
Surprisingly, the variation in the ejecta angle $\Psi_0$ as a function of the geometric ratio $D_{\rm tar}/D_{\rm jet}$ in the granular jet system quantitatively tracks the variation in the water jet near the target~\cite{cheng_2007}, where $D_{\rm tar}$ is the target width and $D_{\rm jet}$ is the jet width. 

In the regime we focus on here, the volume fraction of grains within the jet is close to random close packing, impact creates a dead zone comprised of grains which are effectively immobile directly over the target (see Fig.~\ref{fig:ejected_sheet}{\bf (b)}). Note that because of the high packing fraction, the deadzone is distinct from a shock because the packing fraction remains at random close packing inside and outside of the deadzone.

In an earlier studies~\cite{deadzone, nich_friction}, we found that removing frictional effects at the target eliminates the dead zone but leaves the ejecta dynamics essentially unaffected. Because the existence of the dead zone is not essential to the emergence of the liquid-like ejecta dynamics, we here opt to analyze the idealized problem of impact onto a frictionless target. The absence of a dead zone simplifies the dynamics, allowing us to precisely identify the continuum dynamics controlling the collimated ejecta. Specifically, we compare predictions from an exact solution for the impact of a perfect-fluid jet with results from a full discrete element method simulation. The two sets of results agree quantitatively, demonstrating that dense granular impact simply corresponds to Euler flow at leading order. 

In the rest of the paper, we first describe the idealized impact problem in more detail, then give the numerical results showing that the granular jet during impact remains ``cold,'' where the energy contained in fluctuations is small, and that the large-scale motion produced by impact is to a high degree incompressible and irrotational. These three features suggest that the impact in essence corresponds to the impact of a perfect fluid. Finally we check this idea explicitly by comparing the numerical results of kinematic and stress fields against exact solutions for perfect fluid flow and obtain good agreement. Here we present only results from a 2D impact onto a frictionless target where $D_{\rm jet}$, the width of the granular jet, is equal to the target width $D_{\rm tar}$. Impact in 3D or for different $D_{\rm tar}/D_{\rm jet}$ ratios do not significantly change the results described here. 

Previous experiments used copper and glass beads, both of which are fairly rigid, and found that changing the impact speed does not change the ejecta dynamics, only the characteristic scale of the forces present. In our simulations we therefore idealize the particles within the granular jet as perfectly rigid circular disks. To avoid spurious crystallization in the densely packed jet, we use poly-disperse particles, with diameters uniformly distributed between $0.8D_{\rm grain}$ and $1.2D_{\rm grain}$. Unless otherwise specified, our simulations used $D_{\rm jet} = 200 D_{\rm grain}$, a ratio within the range used in the experiment. We have varied both the coefficient of friction and the coefficient of restitution between grains. We found, in agreement within the range accessed by experiments~\cite{cheng_2007} that their effects are small. The results presented here use a coefficient of friction of $0.2$ and a restitution coefficient of $0.9$. 

To simulate the impact, we use an time step driven scheme modified to handle dense flows efficiently~\cite{nich_sim}. We have compared results from the simulation against the experiment for 3D impact onto a frictional target and verified that we obtain quantitative agreement between the two~\cite{deadzone}.

In 2D, collision with the target creates two collimated ejecta streams (see Fig.~\ref{fig:ejected_sheet}{\bf (c)}). The simulations with a frictionless target also clearly show that the entire interior of the jet is in continual motion, with the velocity varying smoothly from the target to the free surface. To characterize this collective motion, we time average the motions of the individual grains over a time scale $30 D_{\rm tar}/U_0$ spatially bin grains into bins of linear dimension $2.5D_{\rm grain}$ to obtain $\ave{{\v u(\v x)}}$ and $\ave{p(\v x)} = \ave{(\sigma_{xx}+\sigma_{yy})/2}$, velocity and pressure fields that vary spatially. 

\begin{figure}
\centering
\subfigure{\includegraphics[width=0.75\columnwidth]{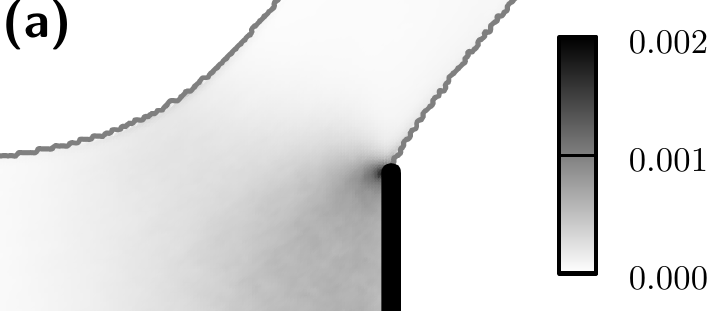}}
\subfigure{\includegraphics[width=0.75\columnwidth]{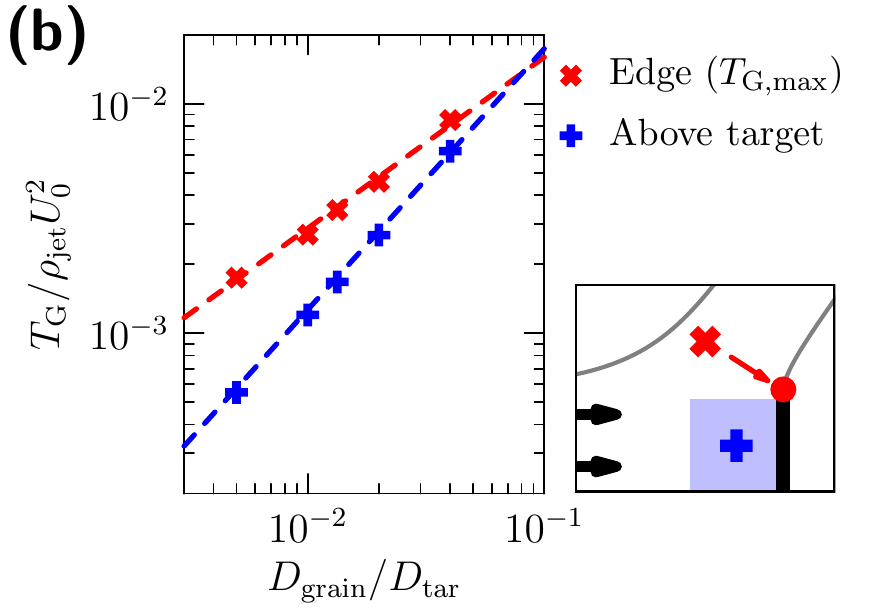}}
\caption{\label{fig:temperature}(Color online) {\bf (a)} Fluctuations in the kinetic energy density, or the ``granular temperature'' $T_{\rm G}$, divided by the impact energy density scale $\rho_{\rm jet} U_0^2$ for half of the jet near the target. The granular temperature $T_{\rm G}$ is about $10^{-3}$ of the impact energy when $D_{\rm tar}/D_{\rm grain} = 200$. {\bf (b)} Normalized average granular temperature in the warm region above the target (\textcolor{blue}{\ding{58}}) and in the hottest region at the target edge (\textcolor{red}{\ding{54}}) plotted against the relative grain size $D_{\rm grain}/D_{\rm tar}$. By decreasing the grain size $D_{\rm grain}/D_{\rm tar}$, we suppress fluctuations within the jet.}
\end{figure}


We can gauge whether the collimated flow in granular jet impact is due to the presence of equilibrium hydrodynamics by considering the energy fluctuations within the jet. For example, in the case of liquid jet impact under the laboratory conditions in~\cite{clanet_2001}, the energy density of fluctuations $N k_{\rm B} T/V$ normalized by the impact energy density scale $\rho_{\rm jet} U_0^2$ are about $10^3$. In this case, we know that the water is near thermodynamic equilibrium and the macroscopic flow is thought of as a small perturbation to an equilibrium liquid phase. The equilibrium status for granular flow is quite different from this, as we show here. To assess the relative importance of energy fluctuations in dense granular jet impact, in Fig.~\ref{fig:temperature}{\bf (a)} we plot the amplitude of the fluctuations in the energy density within the jet $T_{\rm G} = \rho_{\rm jet}(\langle u^2 \rangle - \ave{u}^2 )/ 2$, known as the ``granular temperature,'' normalized by the impact energy density scale $\rho_{\rm jet} U_0^2$. 
The normalized amplitude of the energy fluctuations normalized by the impact energy scale is about $10^{-3}$ when $D_{\rm tar}/D_{\rm grain} = 200$. In Fig.~\ref{fig:temperature}{\bf (b)} we see that as the grain size $D_{\rm grain}/D_{\rm tar}$ is decreased, the granular temperature decreases. In the continuum limit of infinitesimally small grains, the energy fluctuations appear to vanish. We conclude that the granular jet is far from equilibrium, and that the continuum flow, and therefore also the collimated ejecta, is not the result of an equilibrium hydrodynamical phase on impact.

\begin{figure}
\centering
\subfigure{\includegraphics[width=0.75\columnwidth]{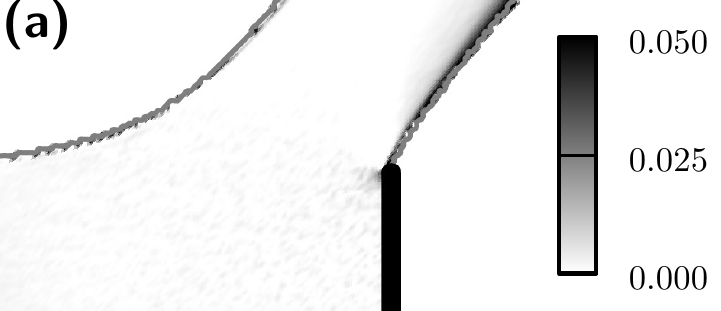}}
\subfigure{\includegraphics[width=0.75\columnwidth]{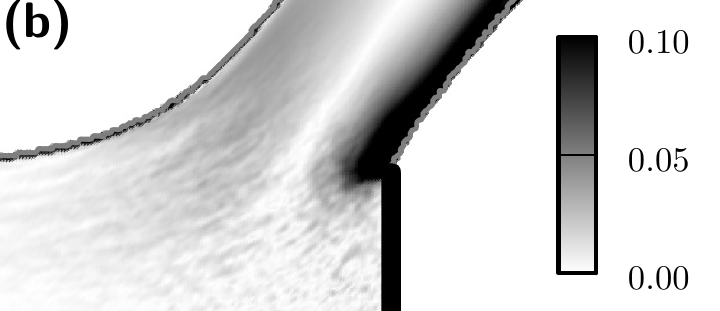}}
\caption{\label{fig:divCurl} We plot {\bf (a)} the nondimensionalized divergence $D_{\rm tar}(\boldsymbol\nabla\cdot\ave{\v u})/U_0$ and {\bf (b)} vorticity $D_{\rm tar}|\boldsymbol\nabla\wedge\ave{\v u}|/U_0$ for half of the jet. The divergence and more noticeably the vorticity become large on the free streams behind the target, in the regime of ballistic motion where grain-grain interactions are weak. But near the target, where the collimated response is produced, both the vorticity and divergence remain small.}
\end{figure}

In Fig.~\ref{fig:divCurl}{\bf (a)} we show the non-dimensionalized divergence $|\boldsymbol\nabla \cdot \ave{\bf u}| / (U_0 / D_{\rm tar})$ throughout the jet. We find that the divergence of the mean velocity field is small, on the order of $0.1\%$ of the characteristic scale of the velocity gradient.
In Fig.~\ref{fig:divCurl}{\bf (b)} we examine the non-dimensionalized vorticity, $|\boldsymbol\nabla \wedge \ave{{\bf u}}| / (U_0 / D_{\rm tar})$. The normalized vorticity in the majority of the jet is, at roughly $1\%$ the magnitude of the velocity gradient created by the impact, relatively small.
The only exceptions to the flow being incompressible and irrotational are regions near the free surface of the ejecta sheet, where the particles have ceased colliding with each other. This suggests that the average motion is, to a high degree, incompressible and irrotational throughout the impact region. 

A perfect fluid describes the flow of an incompressible continuum material uninfluenced by thermodynamics. The vorticity of a perfect fluid does not change in time, hence a perfect fluid in jet impact will be irrotational before impact and remain irrotational. The finding that granular jet impact is largely incompressible and irrotational suggests that the continuum limit for dense granular impact can be idealized as a perfect fluid.

The impact of a perfect fluid satisfies the Euler equations 
\begin{equation}\label{eq:euler}
\rho_{\rm jet} (\del_t + \v u\cdot\boldsymbol\nabla)\v u = - \nabla p, \qquad \boldsymbol\nabla \cdot {\bf u} = 0
\end{equation}
everywhere in the jet interior. The surface shape of the jet is not known {\em a priori}, but is determined along with the rest of the flow when subject to the following boundary conditions. The pressure remains constant along the free surface of jet. Far upstream and downstream of the target, the velocity within the jet is uniformally translating. In 2D and in steady state, an exact solution describing impact of a perfect fluid jet can be obtained via classical conformal mapping techniques~\cite{milne-thomson, gurevich}. 

Perfect fluid flow conserves energy, yet the granular jet has dissipation by means of inelasticity and Coulombic friction. For this reason, we expect and indeed find some discrepancy between the Euler flow solution and the granular jet simulation results. Specifically, because of these dissipative processes the velocity in the ejecta sheet does not recover the incident impact speed $U_0$, as it should if energy is conserved. Instead the exit speed is reduced slightly from the incident value. Since the flow is incompressible, this causes the ejecta sheet in the granular simulation to be slightly wider than the ejecta sheet created by perfect fluid impact. The same reason causes the ejecta angle to be slightly lower for the granular jet impact ($\Psi_0 = 51^\circ$) than that of a perfect fluid ($\Psi_0 = 54^\circ$). 

\begin{figure}
\centering
\subfigure{\includegraphics[width=1.0\columnwidth]{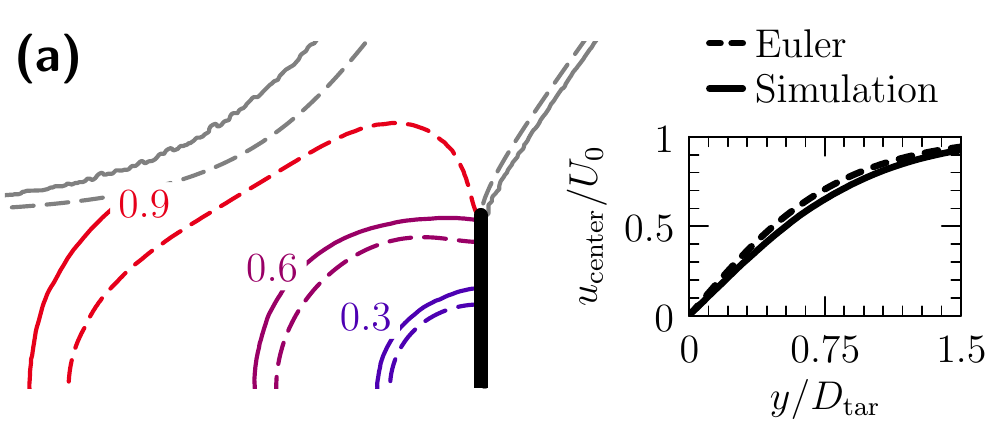}}
\subfigure{\includegraphics[width=1.0\columnwidth]{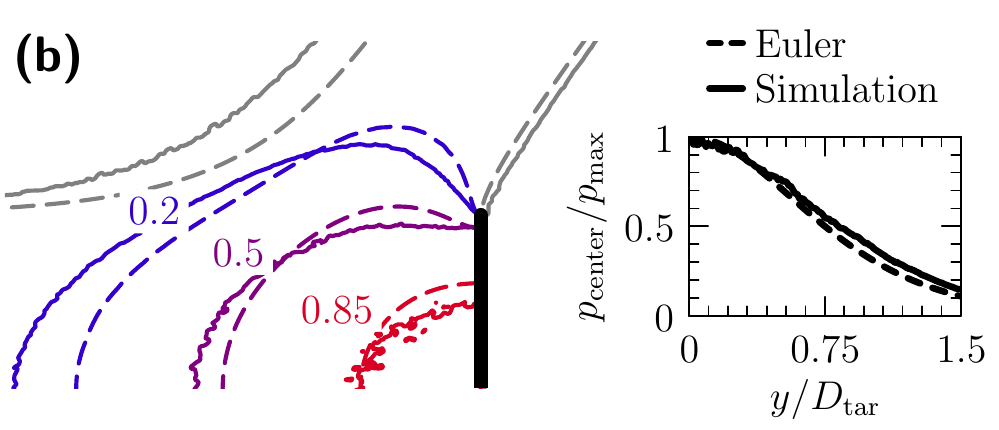}}
\caption{\label{fig:data} (Color online)
Comparison of {\bf (a)} the normalized speed field $|\v u(\v x)|/U_0$ and {\bf (b)} the normalized pressure field $p(\v x)/p_{\rm max}$ between an exact solution for 2D perfect fluid jet impact (dashed lines) against a simulated 2D jet impact (solid lines) for half of the jet. The maximum pressure $p_{\rm max}$ is achieved at the target center. Values from the simulation are time averaged. Here $D_{\rm tar}/D_{\rm jet} = 1$. The $y$ coordinate is the distance above the target along the direction of incoming jet (Fig.~\ref{fig:ejected_sheet}{\bf (c)}). Euler flow captures the speed and pressure fields of the granular jet to leading order.} 
\end{figure}

In Fig.~\ref{fig:data}{\bf (a)} we plot speed contours predicted by the exact solution against numerical results at $3$ different contours, as well as a comparison along the center line of the jet. We normalize the speed by dividing the local velocity by the impact speed $U_0$. The largest discrepancy occurs along the boundary of the jet, as evidenced by the $0.9$ speed contour intersecting the jet surface profile from the numerics. This is because the velocity along the surface of granular jet slows slightly due to energy dissipation. Near the target, the contours agree quantitatively. Overall the agreement is surprisingly good. 

Having found that the Euler flow solution reproduces, at leading order, the kinematics of dense granular impact, we next examine its prediction about the stress distribution within the jet. Since Euler flow is purely driven by gradient of the pressure field, i.e. it supports $0$ tangential stresses, in Fig.~\ref{fig:data}{\bf (b)} we compare the pressure from Euler flow against the pressure from the granular impact simulation. We normalize the pressure by the maximum pressure $p_{\rm max}$ achieved at the center of the target. The pressure data from numerics agrees quantitatively with predictions from perfect fluid flow. 
However, while the speed contours exhibit little fluctuation throughout the jet interior, the pressure contours show larger and larger fluctuations as the target is approached. 
This difference arises because the transmission of forces within the granular jet occurs via force chains. A small positional rearrangement of the grains can cause a large rearrangement of how stresses distribute through the jet.
Aside from these fluctuations, we nevertheless see good agreement. Granular jet impact, at leading order, is controlled by the same pressure profile as that governing the impact of a perfect fluid jet. 

The finding that dense granular impact onto a frictionless target is controlled, at leading order, by a particularly simple continuum behavior raises an open question: what are the necessary ingredients for impact to produce highly collimated ejecta?

Another example of an unexpected emergence of a liquid-like response in a system of particles after impact comes from the relativistic heavy ion collider (RHIC).
RHIC was designed to produce quark gluon plasma, a state of matter at a very high energy density obtainable only by means of heavy nucleon collisions.
Collisions at RHIC did not produce an isotropic scattering pattern, as is found in collider experiments using lighter hadronic particles. Instead the ejecta is collimated as evidenced by elliptical momentum distribution in the ejecta. This collimation has been interpreted as a signature that a liquid quark-gluon phase was created by the impact~\cite{rhic_jacak,rhic_romatschke}. However, the presence of local equilibrium throughout the entire beam during impact at RHIC is currently not fully justified~\cite{song_rhic}, and therefore coherent ejecta at RHIC might not be due to the material phase on impact.

In dense granular jet impact we also find difficulty attributing collimation to a liquid phase, since the jet is far from equilibrium. This finding together with the fact that the presence or absence of a dead zone in the jet interior introduces only minor changes in the ejecta dynamics suggests to us that impact generically induces a liquid-like ejecta response when the initial particle density is sufficiently large. Collimation occurs regardless of what happens to the material within the high-pressure region. We have provided supporting evidence for this view by identifying the continuum limit controlling the ejecta formation as perfect fluid flow. In other words, the observed liquid-like response can be quantitatively accounted for by considering the mechanics of continuum impact alone. It is not necessary to invoke the existence of a liquid phase.

Our results also shed light on important technological processes, such as blast cleaning, in which a stream of glass beads are accelerated towards engine blades in order to clean them. More broadly, the discovery that impact onto a frictionless target simplifies dense granular flow into that of a perfect fluid allows us probe more realistic impact conditions analytically, via perturbative expansions, thereby potentially providing a more rigorous understanding of how dense granular flow organizes during impact. 

In conclusion, we demonstrate here that the impact of a dense granular jet onto a frictionless target is controlled by Euler flow. Predictions from an exact solution for perfect fluid impact quantitatively reproduce the velocity and pressure fields found numerically. Discrepancies between the two solutions can be understood largely as a consequence of the fact that the perfect fluid impact solution conserves energy while dissipative processes, such as inelasticity and friction, are present within the granular jet. This correspondence is important because it demonstrates that a liquid-like dynamical response, the ejection of a collimated sheet after impact, owes its existence to incompressibility and momentum conservation, i.e. Eq.~\eqref{eq:euler}. The existence of a liquid phase is unnecessary.

\begin{acknowledgments}
The authors thank S.~R.~Nagel and H.~Turlier for their experimental results on granular jet impact, as well as X.~Cheng, E.~Efrati, H.~M.~Jaeger and M.~Miskin for helpful discussions. This work was supported by a DOEd GAANN Fellowship (J.E.), NSF MRSEC DMR-0820054 (N.G.), and NSF CBET-0967282 (W.W.Z.).
\end{acknowledgments}
\bibliography{paperBib}

\end{document}